\documentclass[12pt]{article}

\usepackage{amsmath,amsthm,amssymb}
\newcommand{\bonus}{\mathfrak{B}}
\newcommand{\gred}{\textbf{Red}\,}
\newcommand{\ggreen}{\textbf{Green}\,}
\newcommand{\gleft}{\textbf{Left}\,}
\newcommand{\gright}{\textbf{Right}\,}

\title{Memento Ludi: Information Retrieval from a Game-Theoretic Perspective}

\author{George Parfionov\thanks{Friedmann Laboratory For Theoretical Physics,
Department of Mathematics, SPb EF University, Griboyedova 30--32,
191023 St.Petersburg, Russia} and Rom\`an Zapatrin\thanks{Department
of Information Science, The State Russian Museum, In\.zenernaya 4,
191186, St.Petersburg, Russia (corresponding author, e-mail
zapatrin@rusmuseum.ru)}}

\begin{document}

\maketitle

\begin{abstract}
We develop a macro-model of information retrieval process using Game
Theory as a mathematical theory of conflicts. We represent the
participants of the Information Retrieval process as a game of two
abstract players. The first player is the `intellectual crowd' of
users of search engines, the second is a community of information
retrieval systems. In order to apply Game Theory, we treat search
log data as Nash equilibrium strategies and solve the inverse
problem of finding appropriate payoff functions. For that, we
suggest a particular model, which we call Alpha model. Within this
model, we suggest a method, called shifting, which makes it possible
to partially control the behavior of massive users.

This Note is addressed to researchers in both game theory
(providing a new class of real life problems) and information
retrieval, for whom we present new techniques to control the IR
environment.
\end{abstract}

\section*{Introduction}\label{sintro}

The techniques we present are inspired by the success of
macro-approach in both natural and social science. In
thermodynamics, starting from a chaotic motion of billions of
billions of microparticles, we arrive a simple transparent
strongly predictive theory with few macro-variables, such as
temperature, pressure, and so on. In models of market behavior the
chaotic motion is present as well, but there are two definite
parties, each consisting of a big number of individuals with
common interests, whose behavior is not concorded.

\medskip

From a global perspective, information retrieval looks similar:
there are many individual seekers of knowledge, on one side, and a
number of knowledge providers, on the other: each are both chaotic
and non-concorded. There are two definite parties, whose members
have similar interests, and every member of each party tends to
maximally fulfill his own interests. How could a Mathematician help
them? At first sight, each party could be suggested to solve a
profit \emph{maximization} problem. But back in 1928 it was J. von
Neumann who realized this approach to be inadequate: you can not
maximize the value you do not know \cite{neumann28}. In fact, the
profit gained by each agent depends \emph{not only on its} actions,
but also on the activities of its counterpart, which are not known.
Then the game theory was developed replacing the notion of
optimality by that of acceptability. Similarly, the crucial point of
information retrieval, in contrast to data retrieval, is to get some
satisfaction (feeling of relevance) rather than retrieve something
exact. The analogy

\medskip

{\scriptsize
%\begin{table}[h!]\label{tanalogy}
\begin{tabular}{ccc}
  %\hline
  \begin{tabular}{|ccc|}
  \hline
  \textbf{Data Retrieval} &$\longrightarrow$& matching \\
  \hline
  \textbf{Information Retrieval} &$\longrightarrow$& relevance \\
  \hline
\end{tabular}
 &$\simeq$&
 \begin{tabular}{|ccc|}
  \hline
  \textbf{Optimization} &$\longrightarrow$& maximum \\
  \hline
  \textbf{Game Theory} &$\longrightarrow$& equilibrium \\
  \hline
\end{tabular}
 \\
  %\hline
\end{tabular}
%\end{table}
}%small
\medskip

\noindent was a starting point for us to explore applications of
game theory to the problems of information retrieval.

\medskip

The standard problem of game theory is seeking for reasonable (in
various senses) strategies. When the rules of the game are given,
there is a vast machinery, which makes it possible to calculate such
strategies. In information retrieval we have two parties whose
interaction is of exactly game nature, but the rules of this game
are not explicitly formulated. However, we may observe the
consequence of these rules as users behavior, that is, we deal with
the inverse problem of game theory, studied by Dragan \cite{dragan}
for cooperative games. In this Note we expand it to non-cooperative
case.It turns out that the solution of the inverse problem is
essentially non-unique: different rules can produce the same
behavior. We suggest a particular class of models, called Alpha
models describing an idealized search system similar to Wolfram
Alpha engine.

\medskip

What can search engine managers benefit of these techniques? Game
theory can work out definite recommendations how to control the
interaction between the parties of the information retrieval
process. This sounds unrealistic: can one control massive chaotic
behavior? Thermodynamics shows us that the answer is yes. We can
not control individual molecules, but in order to alter their
collective behavior we are able to change macroparameters: the
engine of your car reminds it to you. In our case the payoff
functions of the Alpha model are just those parameters.

\medskip

In Section \ref{sgamebas} we introduce (only the  necessary) basic
notion from game theory, in Section \ref{smacroir} we formulate the
information retrieval process in terms of game theory and formulate
our method as the inverse problem in game theory. In Section
\ref{salpha} we suggest its particular solution, which we call Alpha
model as it resembles Wolfram Alpha engine and in Section
\ref{smagic} we suggest a method to control massive users' behavior.

\section{Direct problem: classical game theory}\label{sgamebas}

Game theory is a mathematical theory studying conflicts and
trade-offs. It involves rational participants who follow formal
rules. A game is specified by its players, players' strategies and
players' payoffs. Begin with a well-known example (a reformulated
Prisonners' dilemma \cite{tucker}).

There are two players $A$ and $B$. The player $A$ can choose color:
\gred or \ggreen, while $B$ chooses direction: \gleft or \gright.
The rules of the game are specified by the following pair of
\emph{payoff matrices} (Table 1)

\medskip

\begin{table}[h!]\label{tdomination}
\begin{center}
\begin{tabular}{ccc}
\begin{tabular}{|p{14mm}|p{10mm}|p{10mm}|}
\hline
&\gleft&\gright\\
\hline\gred&10&25\\
\hline\ggreen&5&20\\
\hline
\end{tabular}
&\qquad\qquad&\begin{tabular}{|p{14mm}|p{10mm}|p{10mm}|} \hline
&\gleft&\gright\\
\hline\gred&11&4\\
\hline\ggreen&23&17\\
\hline
\end{tabular}\\
&&\\ The gain of $A$&&The gain of $B$
\end{tabular}
\caption{A game with domination, defined by its pair of payoff
matrices having the following meaning: if $A$ chooses \ggreen and
$B$ chooses \gright, $A$ gains 20 and $B$ gains 17, and so on.}
\end{center}
\end{table}

\noindent The Mathematician can predict the outcome of this game
provided the players are rational, namely, wishing to gain more: the
rational player $A$ will necessarily choose \gred and $B$ will
choose \gleft.

\medskip

However, both players know the payoff matrices, so, being rational,
why can't they \emph{agree} for $A$ to choose \ggreen and for $B$ to
choose \gright? The point is that they are acting independently,
which exclude any agreement. This kind of games are called
\emph{non-cooperative} and this is the case for the IR community.

\bigskip

The peculiarity of the above mentioned example is that it has a
unique (and therefore straightforward) solution. However, such kind
of examples does not describe the generic situation. Now let us
consider a more general example (Table 2).

\medskip

\begin{table}[h!]
\label{example2eq}
\begin{center}
\begin{tabular}{ccc}
\begin{tabular}{|p{14mm}|p{10mm}|p{10mm}|}
\hline
&\gleft&\gright\\
\hline\gred&10&20\\
\hline\ggreen&5&25\\
\hline
\end{tabular}
&\qquad\qquad&\begin{tabular}{|p{14mm}|p{10mm}|p{10mm}|} \hline
&\gleft&\gright\\
\hline\gred&11&4\\
\hline\ggreen&17&23\\
\hline
\end{tabular}\\
&&\\ The gain of $A$&&The gain of $B$
\end{tabular}
\caption{A non-dominating case: two Nash equilibria.}
\end{center}
\end{table}

First note that no player has a dominating strategy here, so the
outcome of the game is at first glance unpredictable. However the
Mathematician predicts us the outcome of this game as well. First,
we see that both (\ggreen, \gleft) and (\gred, \gright) will
not\footnote{How it works: suppose $A$ chooses \ggreen, observes
that he gains only 5 and then switches to \gred, which brings him
10.} be realized by rational players. One of the following two pairs
(just according to the maritime Rules of the Road) will necessary
occur: (\gred, \gleft) or (\ggreen, \gright). Why so? The motivation
for a rational player to be abide of certain strategy is that
leaving it \emph{unilaterally} reduces his gain:
\begin{equation}\label{enash0}
   \left\lbrace
   \begin{array}{ccc}
     H_A(\mbox{\gred},\mbox{\gleft}) & \geqslant & H_A(\alpha,\mbox{\gleft}) \\
     H_B(\mbox{\gred},\mbox{\gleft}) & \geqslant & H_B(\mbox{\gleft},\beta)  \\
   \end{array}
   \right.
\end{equation}
where $H_A(\alpha,\beta)$ ($H_B((\alpha,\beta)$, resp.) is the gain
of $A$ ($B$, resp.) when $A$ chooses strategy $\alpha$ and $B$
chooses $\beta$. The relations \eqref{enash0} are the famous Nash
inequalities. A pair of strategies is said to form the \emph{Nash
equilibrium}, if they satisfy these inequalities. In the above
example the pair of strategies (\gred,\gleft) is Nash equilibrium,
but so is the pair (\ggreen,\gright) as well! So, what will be the
Mathematician's prediction for the outcome of this game? He will
point out what will not occur and what will take place stably.

\bigskip

Now let us pass to the next example (Table 3), which is generic.
\medskip

\begin{table}[h!]
\label{examplenonash}
\begin{center}
\begin{tabular}{ccc}
\begin{tabular}{|p{14mm}|p{10mm}|p{10mm}|}
\hline
&\gleft&\gright\\
\hline\gred&10&20\\
\hline\ggreen&5&25\\
\hline
\end{tabular}
&\qquad\qquad&\begin{tabular}{|p{14mm}|p{10mm}|p{10mm}|} \hline
&\gleft&\gright\\
\hline\gred&4&11\\
\hline\ggreen&23&17\\
\hline
\end{tabular}\\
&&\\ The gain of $A$&&The gain of $B$
\end{tabular}
\caption{No Nash equilibria.}
\end{center}
\end{table}
We see that there is no equilibrium pairs of strategies in this
game, that is, if the players are represented by individuals, the
outcome of an instance of the game can not be predicted. What can
the Mathematician tell us now? He will suggest to consider players
represented by communities. A choice of the strategy by the
collective player $A$ is described by the distribution of the
individuals with respect to the strategies they choose:
\begin{equation}\label{ecollectstrat}
\left\lbrace
\begin{array}{ccc}
    \vec{p} &=& \left(p_\gred,\: p_\ggreen\right)\\
    \vec{q} &=& \left(q_\gleft,\: q_\gright\right)
\end{array}
\right.
\end{equation}
The gain of the collective players with respect to the chosen pair
of strategies is the average:
\begin{equation}\label{enash1}
   \left\lbrace
   \begin{array}{ccc}
     H_A(\vec{p},\vec{q}) &=& \sum a_{jk} p_j q_k \\
     H_B(\vec{p},\vec{q}) &=& \sum b_{jk} p_j q_k  \\
   \end{array}
   \right.
\end{equation}
where $\left[a_{jk}\right]$, $\left[b_{jk}\right]$ are the payoff
matrices for the players $A$ and $B$, respectively.

\bigskip

The prediction of the outcome of the game is now a pair of
distributions $(\vec{p_\ast},\vec{q_\ast})$ obtained from the same
Nash inequalities \eqref{enash1}, but referred now to averages.
\begin{equation}\label{enash2}
   \left\lbrace
   \begin{array}{ccc}
     H_A(\vec{p_\ast},\vec{q_\ast}) & \geqslant & H_A(\vec{p},\vec{q_\ast}) \\
     H_B(\vec{p_\ast},\vec{q_\ast}) & \geqslant & H_B(\vec{p_\ast},\vec{q})  \\
   \end{array}
   \right.
\end{equation}
The fundamental result of game theory is Nash theorem \cite{owen},
which states that the equilibrium in the sense of \eqref{enash2}
always exist. Moreover, when the number of players is two, the
answer can be written explicitly:
\begin{equation}\label{etwoplayerseq}
    \left\lbrace
    \begin{array}{ccccc}
      p_1 & = & \frac{b_{22}-b_{21}}{b_{11}+b_{22}-b_{12}-b_{21}}\vphantom{\int} & ; & p_2=1-p_1 \\
      q_1 & = & \frac{a_{22}-b_{12}}{a_{11}+a_{22}-a_{12}-a_{21}}\vphantom{\int} & ; & q_2=1-q_1
    \end{array}
    \right.
\end{equation}

Note that the behavior of the player $A$ is completely determined
\emph{only} by the payoff matrix of the player $B$ and \emph{vice
versa}.

\section{Crowd Meets Crowd -- Inverse Problem}\label{smacroir}

In this section we describe our IR macromodel as a non-antagonistic
conflict of two parties, or, other words, a cooperative game of two
players. The first player, call it $A$, asks questions, the second,
call it $B$, provides answers. The player $A$ stands for the
community of users (intellectual crowd) of IR systems, the player
$B$ stand for the community of providers of search results (which is
symmetrically treated as intellectual crowd).

Each particular strategy $\alpha_j$ of the player $A$ is just typing
something in a searchbox. Each particular strategy $\beta_k$ of the
player $B$ is to return a page with an answer, which, viewed as,
say, HTML code, is a string of symbols as well. An instance of the
game is a pair
\[
\alpha_j\beta_k=(\mbox{\texttt{input-string},\texttt{returned-string}})
\]
which is somehow evaluated by each participant. For example, the
payoff value $H^A(\alpha_j\beta_k)$ for the player $A$ for the
pair
\[
\alpha_j\beta_k=(\mbox{\texttt{`accommodation'},\texttt{'No
results found'}})
\]
is evidently low. In the meantime we do not dare to ascribe any
payoff value $H^B(\alpha_j\beta_k)$ of this instance for the
player $B$ (we do not know providers' priorities). In more general
situations even the evaluations of the player $A$ is not known as
well.

\medskip

However, \emph{numerical} payoff values are needed in order to apply
game theory: its basic concept
--- of Nash equilibrium --- is based on comparison of instances
\eqref{enash2}. As a matter of fact, the participants of IR process
do compare instances, but they do it qualitatively. But the
Mathematician needs numbers! What data should he proceed in order to
get them?

\paragraph{Stability and equilibrium.} In a sense
this week's World Wide Web is the same as it was a week ago,
whatever be the variety of different queries and answers. What is
\emph{stable in time} is the statistics of instances
\(\alpha_j\beta_k\): things frequently asked yesterday repeat today.
The Mathematician tells us that from a game-theoretic perspective
this stability is not surprising: these are Nash equilibria which
are stable, because leaving them is unfavorable.

If we had known the payoff functions, we could find the Nash
equilibrium. But in our situation we know the equilibrium
(statistics of instances) and we have to find the appropriate
payoff functions $H^A(\alpha_j\beta_k)$, $H^B(\alpha_j\beta_k)$ in
\eqref{enash1}. This is the \emph{inverse problem} in game theory
\cite{dragan}. The inverse problem has multiple solutions: for
given frequencies there are many different payoff matrices
yielding the same equilibrium\footnote{A trivial example of such
non-uniqueness is multiplying the payoff matrix by a positive
number.}. Below, we introduce a specific model, called Alpha model
with the smallest number of free parameters.

\section{Alpha model}\label{salpha}

The raw material for us will be a collection of search strings
with appropriate frequencies and a collection of returned results
with appropriate frequencies as well. According to our model, we
interpret it as realized equilibrium. Now we are about to
reconstruct the payoff functions. First, according to the remark
made above, we assume that the number of different strategies for
both players is the same. If not, we may reach it by appropriate
prepocessing of data, indetifying some data strings.

Note that, given a pair of strategies $(\vec{p},\vec{q})$, there
are (infinitely) many different payoff functions, for which this
pair of strategies is equilibrium. Among all such models, we
consider the simplest one, closest to data retrieval. For this
model, the payoff matrices are diagonal:
\begin{equation}\label{epalpha}
A= \left(
\begin{array}{cccc}
a_1&0&\ldots&0\\
0&a_2&\ldots&0\\
\cdots&\cdots&\cdots&\cdots\\
0&0&\ldots&a_n
\end{array}
\right) \:;\;\qquad B= \left(
\begin{array}{cccc}
b_1&0&\ldots&0\\
0&b_2&\ldots&0\\
\cdots&\cdots&\cdots&\cdots\\
0&0&\ldots&b_n
\end{array}
\right)
\end{equation}
where $a_1,\ldots,a_n;b_1,\ldots,b_n$ are positive numbers.

This feature of this model is that the only valuable answer for
question $\alpha_j$ is $\beta_j$ with the same index $j$, other
answers $\beta_k$ for $k\neq j$  are of zero value. This looks
like Wolfram Alpha search engine, which provides the only answer
to a query, that is why we call our model Alpha.

\medskip

The Nash equilibrium for the game is given by:
\begin{equation}\label{esolution}
    p_j=\frac{b_j^{-1}}{b_1^{-1}+\cdots+b_n^{-1}}\;;\qquad
    q_k=\frac{a_k^{-1}}{a_1^{-1}+\cdots+a_n^{-1}}
\end{equation}
We can check this directly checking Nash inequalities
\eqref{enash2}. It is sufficient \cite{owen} to check it only for
pure strategies
\begin{equation}\label{echeck}
    \le
\end{equation}

Recall that we have the inverse problem, that is, we know
$(\vec{p},\vec{q})$. Its solution is
\begin{equation}\label{egensolution}
    a_j=\frac{a}{q_j}\;;\qquad b_k=\frac{b}{p_k}
\end{equation}
for any fixed positive numbers $a$, $b$. The obtained result shows
us that:
\begin{itemize}
\item The less frequent is a instance, the higher is its value.
\item The value of a question is determined by the frequency of
the reply, and \emph{vice versa}, the value of a reply is
determined by the frequency of the question.
\end{itemize}

\noindent The first statement means that within this model
frequently asked questions have low value for the provider $B$,
and, \emph{vice versa}, rarely delivered answers are of high value
for the user $A$.

\medskip

The magic of Nash theory is captured in the second statement. It
means that the behavior of player $A$ is completely determined
only by the payoff matrix of player $B$. In other words, the
popularity (=frequency) of users' questions depends on priorities
of the answering side rather than on their own priorities.

\section{Shifting of users' behavior}\label{smagic}

So far, we have suggested a quantitative model of IR process. The
aim of this model is not just to describe, but also to give some
means of control to the overall process. There are two parties
involved, each having its own interests. Let us consider what
could the provider $B$ do in order to increase its gain.

At first sight, the strategy $\vec{q}$ should be changed, but the
power of Nash theory is that the answer is immediate: it does not
make sense, any unilateral deviation from the equilibrium is
unfavorable for $B$. The player $B$ can not directly, by ordering,
control the strategy $\vec{p}$ of player $A$, nor its payoff
matrix. So, the only thing $B$ can do is to \emph{change its own
interests}: what remains under control of $B$, is its own payoff
matrix. How it works?

\medskip

A simple suggestion is to multiply all the elements of $B$ by, say,
1957. This suggestion does not affect, as it follows from
\eqref{esolution}, the strategy of player $A$: it is similar to
recalculating your wealth from euro to Italian liras: you may feel
happy, but your wealth will not grow. So far, we have to accept a
normalization condition for the bonuses $b_k$ of $B$ in order to
make them scale-invariant. Let us suppose their total amount
$\bonus$ to be fixed:
\begin{equation}\label{efixbonus}
    \sum_k b_k=\bonus=\mbox{const}
\end{equation}
As it was shown in previous section, the strategy of $A$ depends
\emph{only} on the payoffs of $B$. Hence, changing the matrix $B$
will affect the behavior of its counterpart $A$. Furthermore, the
statistics of instances will change and, therefore, the average
gain of $B$ will change. Let us first calculate how the average
gain $H^B$ of $B$ depends on the parameters of its payoff matrix
\eqref{epalpha}:
\begin{equation}\label{eaverageb}
    H^B(\vec{p},\vec{q})=
    \sum_j b_j p_j q_j
\end{equation}
For any strategies $p_j$, $q_j$. Within our model we know,
however, that in equilibrium $p_j=\frac{b}{b_j}$
\eqref{egensolution}, therefore the \emph{optimal} average gain
is:
\begin{equation}\label{eaverageoptb}
    H^B(\vec{p},\vec{q})=
    \sum_j b q_j = b
\end{equation}
The value of the multiple $b$ can now be derived from
\eqref{egensolution} and the condition $\sum p_k=1$, therefore the
optimal gain of the player $B$ reads:
\begin{equation}\label{ederivb}
    H^B=\left(\sum_j b_j^{-1}\right)^{-1}
\end{equation}

\medskip

Now let us explore how the optimal gain $H^B$ changes under small
variations $d b_k$ of the parameters of the Alpha model. It
follows from the normalization condition \eqref{efixbonus} that
\begin{equation}\label{edbk0}
    \sum \delta b_k=0
\end{equation}
and calculate the gradient of the optimal gain $H^B$:
\begin{equation}\label{egradb} \nabla_k
H^B =
    -\left(\sum_j
    b_j^{-1}\right)^{-2}\cdot\left(-\frac{1}{b_k^{2}}\right)
=b^2\cdot\frac{1}{b_k^{2}}=p_k^2
\end{equation}
The variations $\delta b_k$ are obtained from the gradient $\nabla_k
H^B$ by requiring the conditions \eqref{edbk0} to be satisfied:
\begin{equation}\label{edeltabk}
    \delta b_k=p_k^2-\frac{1}{n}\sum_j p_j^2
\end{equation}
which is unnormalized Yule's characteristic \cite{yule}, reflecting
the diversity of the variety of queries.

\paragraph{The shifting.} Now suppose we are in a position to make
small changes, of the magnitude $\varepsilon$, of the payoff
function of the Alpha Provider. How should we apply them in order to
make the gain of $B$ maximally increase? The answer is given by the
formula \eqref{edeltabk}, according to which the Alpha Provider has
to do the following:
\begin{itemize}
  \item Find out the relative frequencies $p_k$ of users queries
  $\alpha_k$.
  \item Calculate the average of their squares
  $w=\frac{1}{n}\sum p_k^2$
  \item Slightly re-evaluate the instances placing more bonuses on
  queries, whose frequencies are above the threshold value $w$,
  taking them from rarely asked questions, whose frequencies are
  below $w$.
\end{itemize}
As a result, the equilibrium will shift, the frequencies of users'
requests will adjust accordingly and the Alpha Provider will
increase his gain, as it follows from \eqref{eaverageb} by
\begin{equation}\label{ealphagain}
\delta b\;=\;\varepsilon\sum_j \delta b_j \,q_j
\end{equation}

\section*{Conclusions}

So far, we have described the process of Information retrieval as a
non-antagonistic conflict between two parties: Users and providers.
The mathematical model of such conflict is a bimatrix cooperative
game. Starting from the assumption that \emph{de facto} search log
statistics is the Nash equilibrium of certain game, we provide a
method of calculating the parameters \eqref{egensolution} of this
game, thus solving the appropriate inverse problem.

A significant, somewhat counter-intuitive consequence of Nash theory
is that in this class of games the equilibrium, \emph{i.~e.} stable,
behavior of the User is completely determined only by the
distribution of priorities of the Provider. From this, we infer
suggestions for the provider how to affect the behavior of massive
User.

\paragraph{Acknowledgments.} Rom\`{à}n Zapatrin acknowledges support from RBRF, grant (07-06-00119-a).

\end{document}